\documentclass[preprint,showpacs,preprintnumbers,amsmath,amssymb]{revtex4}
\usepackage[draft]{hyperref} 

\usepackage[most]{tcolorbox}
\usepackage{makecell}
\usepackage{booktabs}
\usepackage{tabularx}
\usepackage{mathtools}
\usepackage{dsfont}
\usepackage{mathrsfs}
\usepackage{wrapfig}
\usepackage{xurl}
\usepackage[T1]{fontenc}
\usepackage{pgfplots}

\usepackage{lastpage}

\usepackage{graphicx}
\usepackage{mathrsfs}
\usepackage{bm}
\usepackage{color,graphicx}
\usepackage{graphicx,colordvi}
\usepackage[ngerman,english]{babel}
\usepackage{caption}
\usepackage{subcaption}
\usepackage{setspace}

\newcommand{\bel}[1]{\begin{equation}\label{#1}}

\def\be{\begin{equation}}
\def\ee{\end{equation}}
\def\bea{\begin{eqnarray}}
\def\eea{\end{eqnarray}}
\def\l{\label}

\def\ms{\medskip}
\def\siml{\;\hbox{\kern.1em \lower.7ex \hbox{$\sim$} \kern-1.12em
\raise.5ex \hbox{$<$} \kern.1em}}
\def\simg{\;\hbox{\kern.1em \lower.7ex \hbox{$\sim$} \kern-1.12em
\raise.5ex \hbox{$>$} \kern.1em}}

\def\d{\hbox{d}}

\def\l{\label}

\def\d{\hbox{d}}

\def\siml{\hbox{\kern.1em \lower.6ex \hbox{$\sim$} \kern-1.12em
          \raise.6ex \hbox{$<$} \kern.1em }}
\def\simg{\hbox{\kern.1em \lower.6ex \hbox{$\sim$} \kern-1.12em
          \raise.6ex \hbox{$>$} \kern.1em }}

\ms

\pacs{21.10.Ev, 21.60.Cs, 24.10.Pa}

\begin{document}
\begin{NoHyper}

\pdfoutput=1

\title{MULTICOMPONENT VAN DER WAALS MODEL OF A NUCLEAR FIREBALL IN THE FREEZE-OUT STAGE}

\author{Yaroslav D.~Krivenko-Emetov}
\email{krivemet@ukr.net, y.kryvenko-emetov@kpi.ua}
\affiliation{National Technical University of Ukraine, 03056, Kiev}

\keywords{multicomponent hadron gas, fireball, freeze-out, van der Waals equation, effective nuclear capability, Grand Canonical Ensemble, pressure fluctuation, quark-gluon plasma, experimental data }

\begin{abstract}
A model of a two-component van der Waals gas is proposed to describe the hadronic stages of nuclear fireball evolution during the cooling phase. During the initial stage of hadronization, when mesons are dominant, a two-component meson model ($\pi^0$ and $\pi^+$ mesons) with an effective two-particle interaction potential in the form of a rectangular well is suggested. In the later stages of hadronization, when almost all mesons have decayed, a two-component nucleon model consisting of protons and neutrons is proposed, incorporating the corresponding effective rectangular nucleon potential. The saddle point method has been utilized for analytical computations of the partition function. This approach has facilitated the consistent derivation of analytical expressions for both pressure and density, considering the finite dimensions of the system, as well as analytical expressions for chemical potentials. It is envisaged that the proposed models and resulting equations can be employed for analyzing experimental data related to the quantitative attributes of particle yields of various types in the final state arising from the hadronic stages of nuclear fireball evolution. Additionally, these models can aid in determining the critical parameters of the system during high-energy nucleus-nucleus collisions. It is demonstrated that in the single-component case, the model's results for the baryonic chemical potential correlate with calculations by other authors.

\end{abstract}





\maketitle
\section*{Introduction}
Experimental observations of elliptical flow in non-central collisions of heavy nuclei at high energies provide substantial evidence that a state of quark-gluon plasma appears and thermalization occurs. This phenomenon is associated with the fact that particles collide with each other multiple times. For this state, one can introduce the concept of temperature, viscosity, density, and other thermodynamic quantities that characterize the substance. In these terms, one can describe and study the phenomena that occur during the cooling of a hadron gas formed after a phase transition from the state of a quark-gluon plasma. It is believed that at a critical temperature ($T > 150$ MeV, the so-called Hagedorn temperature), hadrons "melt," and a phase transition of the hadron gas (hadron matter) into the quark-gluon phase occurs. Therefore, a reverse transition from the quark-gluon phase to the hadron phase is also possible. Therefore, in recent decades, statistical models of hadron gas have been actively used to describe the data of the Large Hadron Collider (LHC), the Relativistic Heavy Ion Collider (RHIC), and even earlier to describe the data of the Alternating Gradient Synchrotron (AGS) and the Super Proton Synchrotron (SPS) on the particle yields in relativistic nuclear-nuclear ($A+A$) collisions at high energies \cite{stachel}, \cite{braun}.

The van der Waals (vdW) model, which takes into account hadron-hadron interactions at short distances, is especially useful in this description \cite{cley} - \cite{yeng}. This is due to the fact that considering the effect of repulsion (excluded volumes) prevents an undesirably high density at high temperatures, a problem that arises in ideal gas models \cite{gorenstein}. Additionally, collisions of heavy high-energy ions in the LHC produce a large number of different particle types. The count of these particles is not fixed. Therefore, the formalism of the Grand Canonical Ensemble (GCE) is an adequate mathematical framework for these phenomena. In this case, thermodynamic quantities do not depend on the number of particles but on the chemical potentials. For many years, researchers have proposed and applied different versions of the vdW models. These models have been primarily used to describe experimental data on the number of particles at high energies, where tens or even hundreds of hadrons of different types can be generated. Naturally, this generation process is limited only by the energy of collisions.

Among these models, the model proposed in \cite{gorenstein} should be noted. In this model, phenomenological parameters of the radii of the hard core, $R_{ii}$ and $R_{ij}$, are introduced. These parameters significantly change the number of yielded particles with different types $N_i$ (where $i$ is the particle type) and are mainly confirmed by experimental data. To describe more subtle effects in the dependence of the hadronic gas pressure on density, various authors (e.g., \cite{krivemet}, \cite{gorvov}) proposed the development of this model \cite{gorenstein}. Here, the effects of attraction between hadrons at a large distance have been taken into account, leading to the appearance of a corresponding contribution to the pressure as $P_{attr} \sim - a n^2$ ($n$ is the density). For a multicomponent gas, the parameter $a$, corresponding to attraction, transforms into parameters $a_{ij}$, and the repulsive parameter $b$ transforms into parameters $b_{ij}$. At the same time, the parameters of the effective potential corresponding to attraction and repulsion depend on the effective radii of repulsion $R_i^0$ and attraction $R_i$, as follows: $a_{ij} \sim u^{(ij)}_0 (c_{ij} - b_{ij})$, $b_{ij} = \frac{2}{3}\pi (R^0_{i} + R^0_{j})^3$, $c_{ij} = \frac{2}{3}\pi (R_{i} + R_{j})^3$, where $u^{(ij)}_0$ is the depth of the effective potential well \cite{krivemet}.

However, even this vdW model cannot be properly developed when considering a finite nuclear system. So, in the case of nuclear collisions, a nuclear fireball with dimensions $<r> \sim 7-10$ fm is observed. In a fairly general case, this problem (without considering the effects of reflection from the system wall) has been solved for a two-component system. In this case, the GCE formalism leads to the use of a double sum, which, in turn, can be transformed into a multidimensional integral. This integral can be integrated using the saddle point method in the vicinity of a saddle point with coordinates $N_1^*, N_2^*$ \cite{krivemet}.

Undoubtedly, it would be beneficial to apply this model to the analysis of experimental data obtained from collisions of heavy nuclei at CERN. One version of such a model was concisely presented in \cite{krivemet1}. It was believed that the collision energies were not high enough, and one could limit oneself to only two particle types: protons and neutrons. It was assumed that characteristic temperatures did not exceed the thresholds at which new particles could form. Therefore, the model itself should transparently possess a non-relativistic limit, while adhering to the conservation law of the total number of nucleons without generating new particles (kinetic freeze-out, cessation of elastic collisions).

The successive stages of the evolution of a nuclear fireball are schematically shown in Figure \ref{pic1}. Moving from left to right: the initial stage with two touching ultrarelativistic nuclei; the state of a hot and superdense nuclear system; gluon and quark-pair creation; quark-gluon phase, representing deconfined nuclear matter expanding hydrodynamically; hadronization and chemical freeze-out (inelastic collisions cease); kinetic freeze-out (elastic collisions cease).

\begin{figure}
\begin{center}
\includegraphics[scale=0.5]{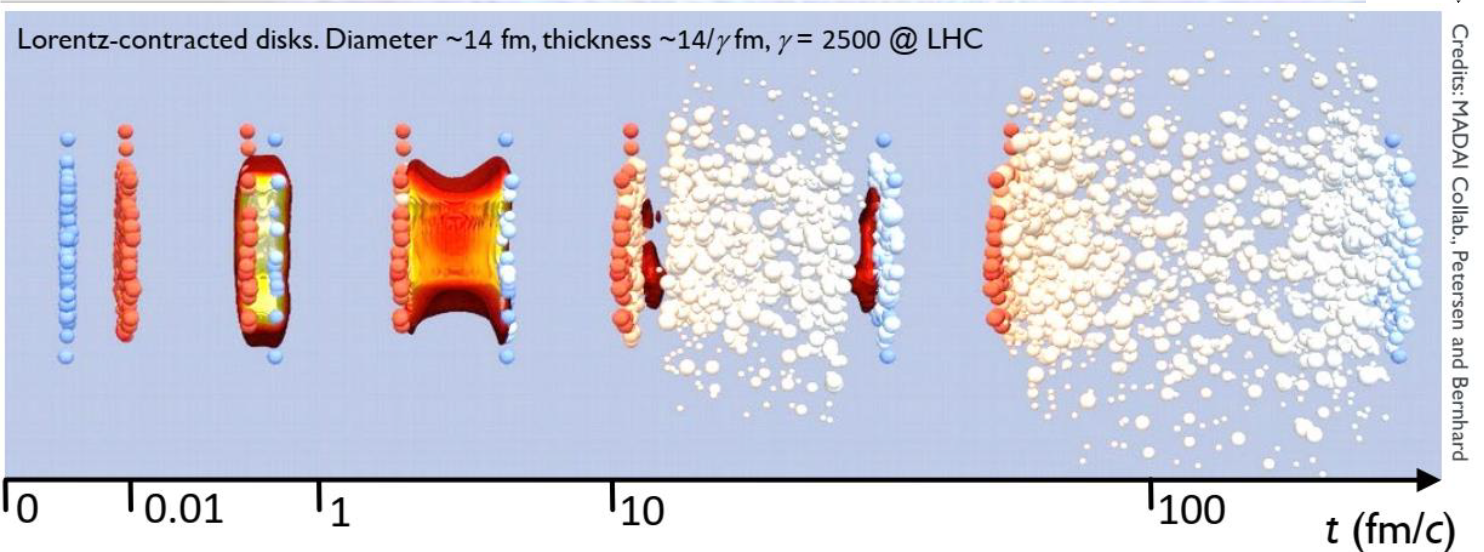}
\caption{Successive stages of nuclear fireball evolution (The figure is taken from \cite{freeze})}
\label{pic1}
\end{center}
\end{figure}

However, at the penultimate stage, after transitioning to the hadronic gas phase, the temperature of the nuclear fireball is approximately $T > 135$ MeV in units where $k_B = 1$ (this corresponds to the stage of hadronization and chemical freeze-out, characterized by inelastic collisions, as depicted in Figure \ref{pic1}).

A more detailed and comprehensive description of the mathematical framework of the model \cite{gorenstein} - \cite{krivemet1} is presented in this article. Some finer effects are estimated, including additional corrections for pressure, density, and root-mean-square (RMS) fluctuations. For situations where temperatures exceed the production threshold ($T > 135$ MeV) and the number of mesons is not conserved, a new two-component meson model \cite{krivemet2} has been proposed.

As the investigation of the hadron fireball and, consequently, the quark-gluon plasma, is expected to be connected with a deeper comprehension of the early universe's evolution, this also underscores the significance of the presented study.

\section{One-component vdW gas}
\label{sec-vdw}
According to various estimates, the duration of the nuclear fireball's existence ($t > 10^{-22}$ s, see Figure \ref{pic1}) exceeds the characteristic time of nuclear interaction $t' \sim 10^{-23}-10^{-24}$ s. This duration of the fireball's existence is compared with the relaxation time $\tau \sim 10^{-21}-10^{-22}$ s for sufficiently small local volumes (subsystems) into which the fireball can be divided.

Therefore, it can be assumed that at each moment in time exceeding the relaxation time, a local statistical equilibrium has had time to establish in the subsystem. In other words, such a local focal area is quasi-stationary, allowing the application of methods from statistical physics. Since all thermodynamic potentials, along with entropy and volume, are positive (extensive) quantities, the corresponding potentials (values) of the entire system (fireball) can be determined as the sum of the corresponding thermodynamic potentials of quasi-closed subsystems \cite{landau}. Accordingly, at each moment in time, a standard representation of the distribution function of a diluted quasi-ideal van der Waals gas in the canonical ensemble (CE) for such quasi-closed subsystems can be provided. In the approximation of pair interactions and under the condition $B(T)N/V \ll 1$, this quantity takes the form \cite{landau}:

\be\l{eq1}
Z(V,T,N)  = \frac{1}{N!} \phi(T,m)^N(V-B(T)N)^N,
\ee
where, respectively, $N$  and  $m$    are the number and mass of particles, $V$   and $T$  are the volume and temperature of the gas.
Formula (\ref{eq1})  uses the notation \cite{gorenstein}:
$$\phi (m,T)=$$

\be\l{eq:14444}
=\frac{1}{2 \pi^2} \int_0^\infty p^2 \exp{\left(-\frac{\sqrt{m^2+p^2}}{T}\right)} dp=\frac{m^2 T}{2\pi^2}K_2(m/T),
\ee
where  $K_2(z)$ is the modified Bessel function, and the second virial coefficient in  (\ref{eq1}) has the form:
\be\l{eq:1555}
B(T)= \frac{1}{2}\int_0^\infty (1-\exp(-U/T)) dV
\ee
and includes pairwise interaction of particles, $U =  U_{ij}$, $(i \neq j)$.

In relativistic limit $ m \gg T $   one can easy obtain, given the asymptotes of the Bessel function:
$$
\phi (m,T) \sim\left(\frac{m T}{2\pi}\right)^{3/2}\exp{\left(-\frac{m}{T}\right)}.
$$
The pressure in the system is easy to find from the partition function:
\be\l{eq2}
\mathcal{P}(V,T,N)= T\frac{\partial}{\partial V}
\ln[Z(V,T,N)]  =  \frac{TN}{V-B(T)N} .
\ee
Note that if the Stirling formula is used in the partition function for the factorial:$$N! \approx \sqrt{2 \pi N}(N/e)^N,$$ then the final pressure formula (\ref{eq2}) will not change.

*MODEL. In accordance with the above, all computations for subsystems will be conducted utilizing methods from statistical physics. This encompasses not only local statistical equilibrium but also the fulfillment of a condition of statistical (thermodynamic) constraint: ($N \rightarrow N_A$), where $N_A$ denotes Avogadro's number.

In such a scenario, the final formulas can be applied to the nuclear fireball due to the mentioned additivity of thermodynamic potentials and volume. Given that the number of particles generated within the fireball reaches around 3-5 thousand during high-energy nuclear interactions, this assumption is reasonably grounded during the initial stages of its evolution.

Of course, at later stages of evolution, this assumption becomes somewhat dubious, as the number of nucleons $N$ within the non-relativistic threshold is constrained by the law of baryon number conservation and equals $N\sim 200$ (heavy element nuclei collide with mass number $A\sim 200$). However, on later stages, the duration of the fireball's existence increases, resulting in an extended relaxation time. Considering these factors, as well as the fact that we can always confine ourselves to the initial stage (see Section 3), it can be considered that the approximation of this model is reasonably justified. It is well-known that practical applications of the van der Waals equation often go beyond the conditions under which the virial approximation was derived, as supported by experience. Therefore, despite the fact that computations in the model are performed using the saddle point method under the condition $B(T)<0$, the final formulas extend to a region where the second virial coefficient $B(T)$ is not necessarily negative.

From the partition function  $Z(V,T,N)$  one can also get: free energy $F(V,T,N)=-T\ln[Z(V,T,N)]$, chemical potential
$$\mu=\left(\frac{\partial F(V,T,N)}{\partial N}\right)=$$
\be\l{eq7771}
=T \left[ \ln (N/V)- \ln (\phi(T,m)) + \frac{ 2B(T) N}{V}\right]
\ee

and the derivative of the chemical potential which in the statistical limit has the form:
$$(  \partial \mu /\partial N ) =-(\partial P/\partial V ) (V/N)^2=$$
\be\l{eq877791}
=  \lim_{N \to N_A} \left( \frac{T}{N}+\frac{ 2B(T)T}{V}\right) \rightarrow \frac{ 2B(T)T}{V}.
\ee
Then, we obtain the Grand partition function (GPF)  $\mathcal{Z}(V,T,\mu)$  from the partition function  $Z(V,T,N)$  taking into account the above physical considerations (see, e.g. \cite{kubo},\cite{feynman}):

\be\l{eq3}
\mathcal{Z}(V,T,\mu)= \sum_{N} \exp\left(\frac{\mu N}{T}\right)Z(V,T,N).
\ee
At high temperatures (which, for example, are realized during collisions of heavy ions in the GCE, and  $ \bigtriangleup N/T \rightarrow  dN'$ ) one can turn from the sum to the integral using the Euler-Maclaurin formula. In this case, the first integral term remains and the logarithm of the statistical sum is introduced into the exponent. Let's denote this indicator by $\Phi(N')$ :
\\
$\mathcal{Z}(V,T,\mu) = T\int_{0}^{ \infty} \d N' 
\exp\left(\mu N'+ \ln[Z(V,N'(T))]\right)=$
\be\l{eq8}
=T\int_{0}^{ \infty} \d N' 
\exp\left( \Phi(N')\right).
\ee


Further integration is performed by the saddle point method \cite{fedor}, since at high temperatures the integrand has a strongly pronounced maximum. We obtain the following expression for finding the maximum point ( $N^{\ast}$ ) for the integrand from the extremum condition imposed on the saddle point:
\begin{scriptsize}
\be\l{eq5}
\mu^*(N^*)=- \left(T\frac{\partial}{\partial N} \ln [Z(V,T,N)]\right)_{N=N^{\ast}} -N^*(\partial \mu /\partial N )_{N=N^*}\approx
\ee
\end{scriptsize}
\be\l{eq51}
\approx T \left[ \ln (N^*/V)- \ln (\phi(T,m) )\right] ,
\ee
where $\mu^*$   is the chemical potential at the saddle point.

As a result, we obtain:
\newline
$\mathcal{Z}(V,T,\mu) = \sqrt{\frac{\pi}{2( \partial^2 \Phi / \partial N^2)_{N=N^*}}}\frac{ \phi(T,m)^{N^*}V^{N^*} }{ \sqrt{2 \pi N^*}(N^*/e)^{N^{*}}}\times$
\be\l{eq511}
\times \exp{ \left (\frac{\mu^*N^*}{T}\right)}  \left (1-\frac{B(T)N^*}{V}\right)^{N^{*}},
\ee
where the second derivative of the exponent   $\Phi$  at the saddle point is defined as follows:
$$(\partial^2 \Phi / \partial N^2)_{N=N^*}=$$
$$ \frac{N^*}{T} (\partial^2 \mu /\partial N^2 )_{N=N^*}+\frac{2}{T} (\partial \mu /\partial N )_{N=N^*}+  (\partial^2 { \ln Z } / \partial N^2 )_{N=N^*}$$
$$=(N^*/T) (\partial^2 \mu /\partial N^2 )_{N=N^*}+ (1/T)  (\partial \mu / \partial N)_{N=N^*} = $$
$$= -\frac{1}{N^*}+\frac{1}{N^*}+\frac{2B(T)}{V} = \frac{2B(T)}{V}<0,$$
because $(N^*/T)\partial^2\mu /\partial N^2 )_{N=N^*}=-1/N^*$ and $ ( \partial \mu /\partial N )_{N=N^*}=-T ( \partial^2 { \ln Z (V,T, N) } / \partial N^2 )_{N=N^*}$.

The pressure in the GCE is defined as follows in terms of the temperature and the logarithm of the GPF (see, e.g., \cite{kubo}):
\be\l{eq:899}
P(T,\mu)=T \frac{\ln \mathcal{Z}(V,T,\mu) }{V}.
\ee

It's easy to show that pressure (\ref{eq:899}), taking into account (\ref{eq511}) and (\ref{eq7771}), can be rewritten as follows:
$$
P(T,\mu^*)  \approx T\xi [1 - B(T)\xi -\ln {(  \partial^2 \Phi^* /\partial N^2 )_{N=N^* }}/(2V\xi)] 
$$
\be\l{eq888}
  \approx T\xi[1-B(T)\xi -\ln {[B(T)\xi ]}/(2V \xi)],
\ee
where the saddle point,  $\xi =N^*(V, T, \mu^*)/V$, is defined according to (\ref{eq51}) and (\ref{eq7771}) as  $\xi  =\phi (m,T) \exp{(\mu^*(\xi) /T)}$.
The parameter   $\xi$ can be eliminated from Eq.\ (\ref{eq888}) using the definition of density, which in the thermodynamic limit turns into the well-known formula \cite{landau}:
\begin{small}
\be\l{eq1000}
n= \frac{\partial P (T, \mu)}{\partial \mu} = \xi [1-2B(T)\xi] - \frac{1}{2V}  \rightarrow \xi [1-2B(T)\xi] .
\ee
\end{small}
In the thermodynamic limit ($N \rightarrow N_A$,  $V \rightarrow \infty$ ) the chemical potential of the saddle point $\mu^*$   from  Eq.\ (\ref{eq51}) when  $N^*=N/(1-2B(T)N/V)$  turns into the chemical potential $\mu$ ( $\mu^*   \rightarrow \mu$ ), which is determined by the well-known
 thermodynamic equation Eq.\ (\ref{eq7771}). 

Both equations  (Eq. (\ref{eq888}) and Eq. (\ref{eq1000})) in parametric form (the saddle point  $\xi$  acts as a parameter) determine the relationship between pressure $P$ , temperature $T$ , and density $n$ . We obtain the state equation in GCE by excluding explicitly this parameter from the system of Eq. (\ref{eq888}) and Eq. (\ref{eq1000}):
\be\l{eq7777}
P(T,n)  \approx  T n[1 + B(T)n] +dP.
\ee
Of course, the resulting state equation is implicitly a parametric equation, since the saddle point  $\xi$ (and, hence, $n$  ) determines the chemical potential $\mu$ according to Eq.\ (\ref{eq7771}) and Eq.\ (\ref{eq51}), as:
\be\l{eq7671}
n=\phi(T,m)\exp(\mu/T -2B(T)n)
\ee

It's crucial that the resulting formula considers the impact of the finite volume of the system, denoted as $V_s$, on pressure. However, the exact nature of this contribution remains unclear to the author. There's a possibility that this might be an unphysical outcome, which could potentially be mitigated by accounting for subsequent terms in the expansion through the saddle-point method. Nevertheless, until a comprehensive analysis is conducted and a quantitative assessment is performed, we will treat this contribution as genuine. It's important to note that this contribution becomes negligible in the thermodynamic limit, where the distinction between CE and GCE disappears.

\begin{large}
$dP=\lim_{V \to V_{s}} \frac{T}{2V} (1 + B(T)n -ln [B(T)n] )  \rightarrow$
\end{large}
\be\l{eq777}
 dP=-\frac{T ln [B(T)n]}{2V_s}.
\ee
If we disregard the correction obtained from the volume of dP and assume that $B(T)n<<1$, then by making the following substitution in the right-hand side of Eq. (\ref{eq7777}): $(1+B(T)n) \sim \exp(B(T)n)$, taking into account Eq. (\ref{eq7671}), it will become the following:
$$P(T,\mu)  \approx  T \phi(T,m)\exp(\mu/T - B(T)n) =$$
\be\l{eq2777}
= T \phi(T,m)\exp(\mu^{int}/T)=P^{id}(T, \mu^{int}).
\ee
Thus, the equation of state with interaction can be obtained by making the substitution $\mu->\mu^{int}=\mu-B(T)n$ in the equation of state of the ideal gas \cite{gorenstein}, \cite{mag}.
These equations are density functionals, which, according to (\ref{eq7771}), at a fixed chemical potential, are found from the solution of a transcendental equation $n=\phi(T,m)\exp(\mu/T -2B(T)n)$.
Assuming $B(T)n<<1$, this formula can be replaced with (\ref{eq1000}) where, according to (\ref{eq51}), $\xi$ is expressed in terms of $\mu$:

\be\l{eq1015}
n = \phi(T,m)\exp(\mu/T)[1-2B(T)\phi(T,m)\exp(\mu/T)] .
\ee

The RMS fluctuations of pressure and density calculated by known formulas (see, e.g.,\cite{landau} \cite{fed}) give estimates of the found corrections to the corresponding quantities:
\begin{small}
\be\l{eq190011}
<(\bigtriangleup P)^2> = \frac{Tn}{V}(  \partial P /\partial n )_{S}\sim (Tn/V)T[1+2B(T)n],
\ee
\end{small}
\begin{small}
\be\l{eq100011}
<(\bigtriangleup n)^2> =\frac{ T}{(  \partial \mu /\partial N )_{N=N^*} V^2} \sim \frac{1}{nV^3}[1-2B(T)n].
\ee
\end{small}

\section{Two-component vdW gas}
\label{sec-vdw2}
Let's examine the procedure for incorporating excluded volume and attraction in the van der Waals model for a two-component hadron gas consisting of two types of particles labeled as "i" and "j", with $N_1$ and $N_2$ being the quantities of particles of the first and second types. In this scenario, the partition function takes the following form\cite{gorenstein}:
$$Z (V, T, N_1, N_2)=$$
\begin{small}
$$
 =\frac{1}{N_{1}!N_{2}!} \int  \prod_{l=1}^{N_1} \frac{d^3p_l^{(1)}d^3r_l^{(1)}}{(2\pi)^3}\exp{\left(-\frac{\sqrt{(m^{(1)})^2+(p_l^{(1)})^2}}{T}\right)}\times
$$
$ \int  \prod_{k=1}^{N_2} \frac{d^3p_k^{(2)}d^3r_k^{(2)}}{(2\pi)^3} \exp{\left(-\frac{\sqrt{(m^{(2)})^2+(p_k^{(2)})^2}}{T}\right)}\exp\left(-U^{(1,2)}/T\right),
$\\
\end{small}
where $m^{(1)}, N_1 (m^{(2)}, N_2)$ are, respectively, the masses and number of particles of the 1st (2nd) sorts, and the two-particle potential has the following form:

$U ^{(1,2)}  =  \sum \limits_{1\leq m<l\leq N_1}^{N_1} u_{11} \left(  \vert \vec{r}^{(1)}_m-\vec{r}_l^{(1)}  \vert\right)+$\\
\begin{footnotesize}
\be\l{eq12}
+ \sum \limits_{1\leq k<s\leq N_2}^{N_2} u_{22} \left( \vert \vec{r}^{(2)}_k-\vec{r}_s^{(2)} \vert\right) +
\sum \limits_{ m=1}^{N_1} \sum \limits_{ k=1}^{N_2}   u_{12} \left(  \vert \vec{r}^{(1)}_m-\vec{r}_k^{(2)}  \vert\right).
\ee
\end{footnotesize}
After a trivial integration over momenta, this expression takes the following form:
$$Z (V, T, N_1, N_2) =\frac{1}{N_{1}!N_{2}!} [\phi (m^{(1)},T)]^{N_1} [\phi (m^{(2)},T)]^{N_2}\times$$
\be\l{eq13}
 \int  \prod_{l=1}^{N_1} d^3r_l^{(1)}
\times \int  \prod_{k=1}^{N_2} d^3r_k^{(2)}\times\exp\left(-U^{(12)}/T\right),
\ee
where the notation introduced is the same as in the first section.

This expression for the pair-interactions approximation ($ U^{(123)}\ll U^{(12)}$ ) and a weakly ideal gas ($ 2N B/V \ll 1$ ) can be rewritten as follows (see \cite{gorenstein},\cite{krivemet} ):

$ Z(V,T,N_1,N_2) \sim\frac{1}{N_1!N_2!} \phi(m^{(1)},T)^{N_{1}} \phi(m^{(2)},T)^{N_{2}}\times$\\
\be\l{eq:19}
\times (V-B_{11}N_1-\tilde{B}_{21}N_2)^{N_1}(V-B_{22}N_2-\tilde{B}_{12}N_1)^{N_2}
\ee
Here the notation $ \tilde{B}_{ij}=2 \frac{B_{ii}B_{ij}}{B_{ii}+B_{jj}}$has been introduced.
The two-particle partition function  $\mathcal{Z}(V,T,\mu_1,\mu_2)$  in GCE is expressed in terms of the 
two-particle partition function  $Z(V,T,N_1,N_2) $  in CE \cite{landau},  \cite{krivemet}, as
$$
\mathcal{Z}(V,T,\mu_1,\mu_2)= $$
$$ =  \sum_{N_1,N_2} \exp\left(\frac{\mu_1 N_1+\mu_2 N_2}{T}\right)
Z(V,T,N_1,N_2). $$\\
Here, as in the one-dimensional case, when $T \gg U$ and $N_i \rightarrow N_A$, the sum over the number of particles approximately becomes an integral, since $\bigtriangleup N/T \to dN'$:

$$\mathcal{Z}(V,T,\mu_1,\mu_2)=$$
$=T^{2}\int_{0}^{ \infty} \d N'_1 \int_{0}^{ \infty} \d N'_2
\exp\left(\mu_1 N'_1 +\mu_2 N'_2\right) Z(V,T,N'_1, N'_2)$
\bea\label{eq4}
=T^{2}\int_{0}^{ \infty} \d N'_1\int_{0}^{ \infty} \d N'_2\exp\left( \Phi(N'_1, N'_2)\right).
\eea
Integration of (\ref{eq4}) by the saddle point method \cite{fedor}  leads us to the following result:
\begin{small}
$$
\mathcal{Z}\sim  \frac { \pi}{ 2  \sqrt{\Phi''(N'^*_1, N'^*_2)}} \exp\left(\frac{\mu^*_1 N_1^*+\mu^*_2 N_2^*}{T}\right)
Z(V,T,N_1^*,N_2^*),
$$
\end{small}
where the coordinates of the saddle point  $N_i^*$  ($i=1,2$)  are found from the extremum conditions:
$\left(\frac{\partial\Phi(N'_i, N'_j)}{\partial N'_i}\right)=0$, 

$\Phi''(N'^*_1, N'^*_2)$    $= det \begin{vmatrix} 
	c_{11}& c_{12}\\
	c_{21}& c_{22}            
\end{vmatrix},$ 

$c_{ij}=\left(\frac{\partial^2\Phi(N'_i, N'_j)}{\partial N'_i\partial N'_j}\right)\mid_{N'=N'^*}.$

Substituting the value of the partition function into the definition of pressure in the GCE \cite{kubo}, we obtain the following expression \cite{krivemet}:
$$
P(T,\mu_1,\mu_2)=T \frac{\ln \mathcal{Z}(V,T,\mu^*_1,\mu^*_2) }{V}\sim $$
\begin{small}
\be\label{eq:25}
\sim T[\xi_1+\xi_2  -\xi_1^2 B_{11}-\xi_2^2 B_{22} - (\tilde{B}_{12}+\tilde{B}_{21})\xi_1\xi_2 -\frac{\ln {C}}{2V}],
\ee
\end{small}
where
$C=|\xi_1 B_{11}\xi_2 B_{22}- \xi_1 \tilde{B}_{12}\xi_2 \tilde{B}_{21}|$.

Using such a mathematical apparatus, one can organically introduce the law of conservation of chemical potentials. The latter are related to the condition imposed on the integrand when finding the saddle point. In the thermodynamic limit the chemical potential determined by the extremum condition coincides with the definition of the chemical potential itself:
\be\label{eq898}
P(T)\rightarrow T[\xi_1+\xi_2  -\xi_1^2 B_{11}-\xi_2^2 B_{22} - (\tilde{B}_{12}+\tilde{B}_{21})\xi_1\xi_2 ] ,
\ee

$\mu^*_i\rightarrow \mu_i=\left(\frac{\partial F(V,T,N_i,N_j)}{\partial N_i}\right),$

where $F(V,T,N_1,N_2)=-T \ln [Z(V,T,N_1,N_2)] $ is the definition of free energy  (\ref{eq51}).

We get from the definition of density 
\be\label{eq339}
n_i= \partial P (T, \mu_i,\mu_j) /\partial \mu_i \sim \xi_i[1  -(2\xi_i B_{ii}+\xi_j (\tilde{B}_{ij}+\tilde{B}_{ji}))].
\ee

The virial expansion (\ref{eq:25}) can be rewritten, taking into account  Eq. (\ref{eq339}), as a two-component vdW equation in the approximation  $b_{ij}N_i/V \ll 1$  and  $a_{ij}/Tb_{ij} \ll 1$) :
$$P(T,\mu_1,\mu_2)   = \frac{Tn_1}{1-b_{11}n_1-\tilde{b}_{21}n_2}+\frac{Tn_2}{1-b_{22}n_2-\tilde{b}_{12}n_1}$$
\bea\label{eq:33}
 -  n_1 (a_{11} n_1+\tilde{a}_{21} n_2)-n_2 (a_{22} n_2+\tilde{a}_{12} n_1)+dP,
\eea
where $dP$ , according to  Eq. (\ref{eq:25}), takes into account the finite size of the fireball. 

When formula (\ref{eq:33}) was derived, the expression   $
 \tilde{B}_{ij} \approx  \tilde{b}_{ij}- \tilde{a}_{ij}/T$ was used (see, e.g., \cite{krivemet}), and for each type of particles the corresponding parameters of attraction and repulsion were introduced:  $a_{ij}$, $\tilde{a}_{ij}  \approx 2 \gamma a_{ij} a_{ii}/(a_{ii}+a_{jj}) $,   $b_{ij}$, $\tilde{b}_{ij}=2 \frac{b_{ii}b_{ij}}{b_{ii}+b_{jj}}$,  $\gamma$ is a phenomenological parameter reflecting the complexity of the problem. Introducing quantities $\tilde{a}_{ij}$ constrained by the condition
$\tilde{a}_{ij}+\tilde{a}_{ji} =2\gamma a_{ij}$.

\section{Two-Component Asymmetric VdW Model with Non-Conservation of Particle Number}
\label{sec-vdww}

As experiments related to the formation of quark-gluon plasma focus on heavy nucleus collisions ($A+A$) with very high energies exceeding 1 GeV per nucleon, it is assumed that at the initial stages of freeze-out, mesons of different types dominate (chemical freeze-out). Therefore, to describe nuclear interactions during this freeze-out stage beyond the threshold for producing new particles ($T>135$ MeV), a generalization of the van der Waals model is proposed for a medium-sized meson fireball \cite{krivemet2}: $<V_f> \sim (V_f^{min}+V_f^{max})/2 \sim (4/3)\pi <a><b>^2 \sim (4/3)\pi r_o^3 <A>\sim 600-1000$ $ fm^3$. Here, $r_0 =1.1-1.2$ $ fm$, $<a>, <b>$ represent the mean semiaxes of the ellipsoid, and $<A>$ denotes the mass number of nuclei remaining in the fireball after the collision. The model assumes that the fireball is primarily composed of mesons, considering that the number of nucleons is much smaller than the number of mesons ($ N_{pn}\sim 200-300 <<N_{\pi,\rho,\omega} \sim 3000-5000$). Contributions from other particles are neglected in the model. Thus, the following natural assumptions are summarized in the model:

1.) The average energies of nucleon-nucleon interactions do not exceed the threshold for producing heavy mesons. Therefore, the model is limited to two types of particles ("0" corresponds to $\pi^0$-mesons, "+" corresponds to $\pi^+$-mesons).

2.) Since reactions producing $\pi^+$-mesons are more likely than reactions producing $\pi^0$-mesons, it is assumed that $ n_0 = k n_+=n $, where $ k <1 $, $ n_0 $ represents the density of $\pi^0$-mesons, and $ n_+ $ represents the density of $\pi^+$-mesons. For instance, this corresponds to a higher probability of $\pi^+$-meson production in reactions like $p+d \rightarrow d+n+\pi^+$ compared to $\pi^0$-meson production in reactions like $p+d \rightarrow d+p+\pi^0$ (also, the lifetime of $\pi^+$-mesons is longer than that of $\pi^0$-mesons).

3.) An effective phenomenological potential of meson interaction $U^{(1,2)}=U^{(i,j)}$ is introduced, where $(i,j)={+,0}$. "$(0+)"$ denotes the interaction of $\pi^0$-mesons with $\pi^+$-mesons, "$(++)"$ denotes the interaction of $\pi^+$-mesons, and "$(00)"$ denotes the interaction of $\pi^0$-mesons. For a gas composed of multiple components, the parameter denoted as $a$, which signifies attraction, undergoes transformation into distinct parameters $a_{ij}$. Similarly, the repulsive parameter $b$ is replaced by parameters $b_{ij}$. This transformation occurs concurrently with the dependence of the effective potential's attraction and repulsion parameters on the effective radii associated with repulsion ($R_i^0$) and attraction ($R_i$). Specifically, the relationship can be expressed as follows: $a_{ij} \sim u^{(ij)}_0 (c_{ij} - b_{ij})$, where $b_{ij}$ is calculated as $\frac{2}{3}\pi (R^0_{i} + R^0_{j})^3$ and $c_{ij}$ is obtained as $\frac{2}{3}\pi (R_{i} + R_{j})^3$. Here, $\gamma u^{(ij)}_0$ represents the depth of the effective potential well.
 
\begin{equation}
U^{(i,j)}  =
\left\{
\renewcommand{\arraystretch}{1.2}
\settowidth{\dimen0}{$0$}
\begin{array}{r@{}>{{}}l@{}>{{}}l}
 \infty &   \qquad \; \textrm{if } r< R^0_{i}+R^0_{j},\\
-\gamma u_{0}^{(i,j)} &  \qquad \; \textrm{if } R^0_{i}+R^0_{j}\le r< R_{i}+R_{j},\\
0  &\qquad \; \textrm{if }R_{i}+R_{j} \geq r.
\end{array}
\right.
\label{eq:15}
\end{equation}
As the parameters of the scalar component of the effective phenomenological rectangular well potential are chosen in such a way as to approximately yield the same pressure and density values as the effective mesonic potential (see Figure  \ref{pic2}, where, for instance, the interaction of  $\pi^0$-mesons with $\pi^+$-meson corresponds to  $U^{(+,0)}$), the effective mesonic potential (a) can be substituted with a similar effective phenomenological rectangular well potential (b).

\begin{figure}
\begin{center}
\includegraphics[scale=0.7]{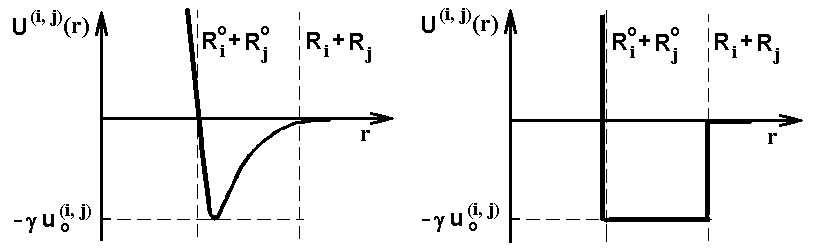}
\caption{ Scalar part of the effective phenomenological meson-meson potential}
\label{pic2}
\end{center}
\end{figure}

4.) It is assumed that the hard-core radius of the $\pi^0$-meson is much smaller than the hard-core radius of the $\pi^+$-meson: $R^0_0 \ll R^0_+$. The hard-core radius of the $\pi^+$-meson is considered to be known.

Average pressure and density fluctuations are easily found within the framework of the proposed model, similarly to formulas Eq. (\ref{eq190011}) and Eq. (\ref{eq100011}):
\be\l{eq100012}
\sqrt{<(\bigtriangleup n)^2>} \sim \frac{1}{\sqrt{nV_f}V_f}[1-B(T)n],
\ee

\be\l{eq190013}
\sqrt{<(\bigtriangleup P)^2>} \sim T\sqrt {n/V_f}[1+B(T)n].
\ee

The following results are obtained (Fig. \ref{pic3}, Fig. \ref{pic4}). Such data have been used (Fig. \ref{pic3}): $T=147$ MeV, the effective radius of the  $\pi^+$-meson, $R^0_+=0.46$ $fm$, and  $\pi^0$-meson, $R^0_0=0.01$ fm, the average value of the volume of the meson fireball is taken as the value $<V_{f}>\sim 600$ $fm^3$,  $k=0.5$, the parameter of the potential depth, $u_{0}^{(+,0)}\sim 80-100$ MeV. One can clearly see (Fig. \ref{pic4}) an increase in the correction $dP/<P>$  at low densities, which is typical in the final stages of the freeze-out.

\begin{figure}
\begin{center}
\includegraphics[scale=0.7]{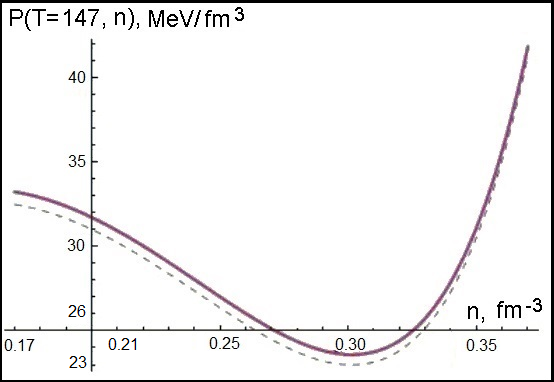}
\caption{ Dependence of nucleon pressure   $P(T, \mu_1,\mu_2, n )$  ( Eq. \ref{eq:33})  on the meson density $ n_0 = k n_+=n $   for the 
two-component asymmetric vdW model with correction (upper isotherm) and without correction (lower isotherm)}
\label{pic3}
\end{center}
\end{figure}

\begin{figure}
\begin{center}
\includegraphics[scale=0.7]{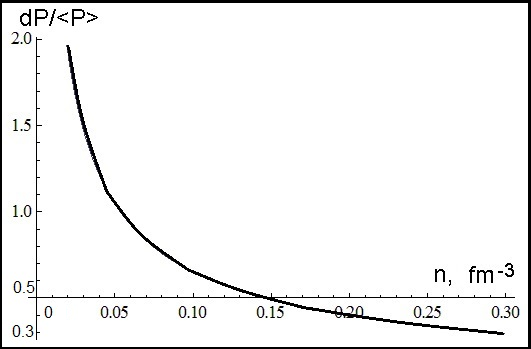}
\caption{ Ratio of the correction to pressure $dP$  from the size of the meson fireball to the value of the RMS pressure fluctuation  $<P>$  as a function of the meson density   $n_0 = k n_+=n $}
\label{pic4}
\end{center}
\end{figure}

\section{Two-Component Asymmetric VdW Model of a Nucleus Fireball at the Final Stage of Freeze-Out}
\label{sec-vdww}

The average lifetime of mesons that dominate in the initial stages of freeze-out is quite short ($\tau \sim 10^{-8}-10^{-16}$ c). In the hadronic medium, due to various reactions, this time is even shorter. Hence, they decay rapidly. Consequently, at the final stage of freeze-out, baryons, particularly protons and neutrons, start to dominate. Additionally, as shown earlier, the effects of finite volume become significant at sufficiently low density values. This formally corresponds to the last stages of the fireball's evolution. Therefore, despite certain doubts about the existence of a fireball at such late stages, when the boundary between the gas and the aggregate of individual nucleons gradually disappears (see the far right part of Figure \ref{pic1}), in order to describe nucleus-nucleus interactions in the last stage of freeze-out, which is below the threshold for producing new particles ($T<135$ MeV) (kinetic freeze-out), the work \cite{krivemet1} proposed the following generalization of the vdW model to the nucleus fireball. By analogy with the previous section, the following simplifications are suggested:

1.) The average energies of inter-nucleon collisions do not exceed the threshold for producing other hadrons. Therefore, for simplicity, the model is limited to two types of baryons ($"p"$ - proton, $"n"$ - neutron).

2.) We use the relation between the densities of protons and neutrons in the form of $n_p + n_n = n$, in accordance with the conservation law of baryon number, $Z + N = A$.

3.) It is assumed that the composition of colliding nuclei is known, such that $n_p = k n_n$, where $k <1$, since heavy nuclei have an excess of neutrons.

4.) The effective potential of proton-neutron, proton-proton, and neutron-neutron interactions, which leads to approximately the same pressure and density values as the real potential (Figure \ref{pic2}), can be represented by analogy to the model in Section 3 as $U^{(i,j)}$, where $i,j=(p,n)$.

5.) The hard-core radius of the proton is considered known, $R^{'0}_p = 0.8$ fm (but in this work, this radius is a phenomenological parameter equal to $R^0_p = 0.5$ fm). It is assumed that the root-mean-square radius of the neutron is much smaller than that of the proton: $R^0_n << R^0_p$(the neutron's magnetic radius is approximately 0.864 fm, but the mean square charge radius is negative and much smaller. This is because the charge structure of the neutron only becomes apparent at non-zero transferred momenta $q$ \cite{atac}).

From the equation Eq.(\ref{eq:33}), it can be derived:
\be\label{eq:334}
 P(T,\mu_1,\mu_2)   = \frac{Tn^*(k+1-\alpha n^*)}{1-\beta n^*+ \delta (n^*)^2} - a^* (n^*)^2 +dP, 
\ee
where 
$dP\cong -T n^*\frac{ ln [C(T,n^*)]}{2A}$, $C(n^*,T)=|n^*B_{11}n^*B_{22}-n^*\tilde{B}_{12}n^*\tilde{B}_{21}|$,
$n^*=n/(1+k)$,  $\alpha=b_{11}k+\tilde b_{12}+\tilde b_{21}k^2+b_{22}k $,\\ $ \beta=b_{11}k+\tilde b_{12}+\tilde b_{21}k+b_{22}$ ,\\$ \delta= k b_{11}b_{22}+k^2 b_{11}\tilde b_{12}+b_{22}\tilde b_{21}+\tilde b_{12}\tilde b_{21}k$,\\  $a^*=a_{11}k^2+(\tilde a_{12}+\tilde a_{21})k+a_{22}$.\\
It follows from the condition $ R ^ 0_2 \ll R ^ 0_1 $  $\Rightarrow$ $ b_{22} \ll b_{11} $, $\alpha\cong \beta$.

Similarly to the equations \ref{eq100012} and \ref{eq190013}, corresponding average fluctuations of pressure and density are determined.

Functional dependences for pressure, obtained by Eq. \ref{eq:33}, and the ratio of  $dP$  to RMS pressure fluctuations are shown in Figs. 5 and 6.

\begin{figure}
\begin{center}
\includegraphics[scale=0.7]{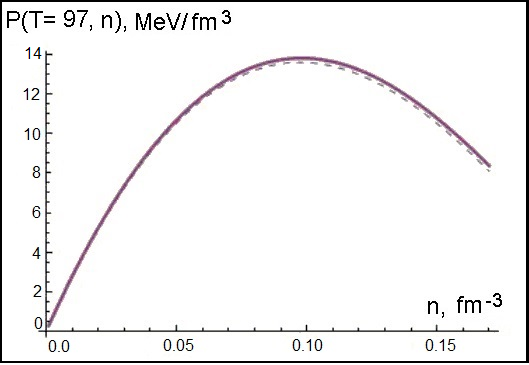}
\caption{ Dependence of nucleon pressure $P$  \ref{eq:334} on nucleon density, $n_p=kn_n=n$ , in the two-component asymmetric vdW model with correction (upper isotherm) and without correction (lower isotherm)}
\label{pic5}
\end{center}
\end{figure}

\begin{figure}
\begin{center}
\includegraphics[scale=0.7]{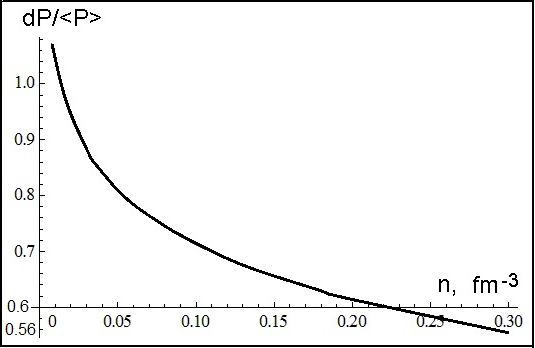}
\caption{ The ratio of correction from the size of the nucleon fireball to pressure $dP$  to the value of the RMS pressure fluctuation  $<P>$ depending on the density of nucleons,  $n_p=kn_n=n$}
\label{pic6}
\end{center}
\end{figure}

It can be seen that the correction  $dP$  makes a nonzero contribution to the total pressure also in this case. On the other hand, it is negligibly small almost everywhere in comparison with the contribution from fluctuations. The correction makes a contribution comparable to fluctuations only in the region near zero density that is nonphysical for a nuclear fireball. But it can be neglected in this region, as can be seen from Fig. 5.

\section{Multicomponent vdW gas}
\label{mult-vdw}

It is possible to extend the above analysis to the vdW gas with multiple components, consisting of any number of different particles. By integrating over the momentum of the particles and making some modifications similar to those done in the first example,
one can obtain an expression for the multicomponent (K-component) vdW gas in the canonical partition function:

$ Z(V,T,N_1,...,N_K) \sim \prod_{p=1}^{K}\frac{1}{N_{{p}}!}\phi(T,m_{p})^{N_{p}}\times$\\
\be\l{eq:1999}
\left (V-\sum \limits_{(p\neq q)=1}^{K} \left( B_{pp}N_p+\tilde{B}_{qp}N_q\right)\right)^{N_{p}}.
\ee

It is possible to calculate the pressure in the Grand Canonical Ensemble for the vdW gas with multiple components by integrating over the namber of the particles and making appropriate modifications to the formula derived for a single-component gas.
$$P(T,\mu_1,...,\mu_K)=\sum \limits_{p=1}^{K}\left[ \frac{Tn_{p}}{1-\sum \limits_{(p\neq q)=1}^{K}(b_{pp}n_p+\tilde{b}_{pq}n_q)} \right]$$
\bea\label{eq:335}
 - \sum \limits_{(p\neq q)=1}^{K}n_p (a_{pp} n_p+\tilde{a}_{qp} n_q)+dP(T,\mu_1,...,\mu_K),
\eea
where $\mu_p=\left(\frac{\partial F(V,T,N_1,...,N_K)}{\partial N_p}\right) $ are the chemical potentials $(p = 1, ...,K)$. 

The particle densities $n_p= \partial P (T, \mu_1,...,\mu_K) /\partial \mu_p$ along with the pressure are obtained as the solutions of the system of related equations depending on the parameter of the saddle points $\xi_p$ $(p = 1, ...,K)$.

The HG model in the grand canonical ensemble formulation does not have fixed numbers for $N_1, ..., N_K$ due to inelastic reactions between the hadrons. However, the conserved charges of baryonic number $B$, strangeness
 $S$ (which is conserved by neglecting weak decays), and electric charge $Q$ have fixed values.
The value of $B$ corresponds to the number of participating nucleons in the reaction, while $S$ is equal to zero. The value of $Q$ is $0.5eA$ for intermediate nuclei and $0.4eA$ for heavy nuclei($Q=eA/(2+0.014A^{2/3})$).
The use of the grand canonical formulation is more advantageous in case of high temperatures.
\begin{figure}
\begin{center}
\includegraphics[scale=0.3]{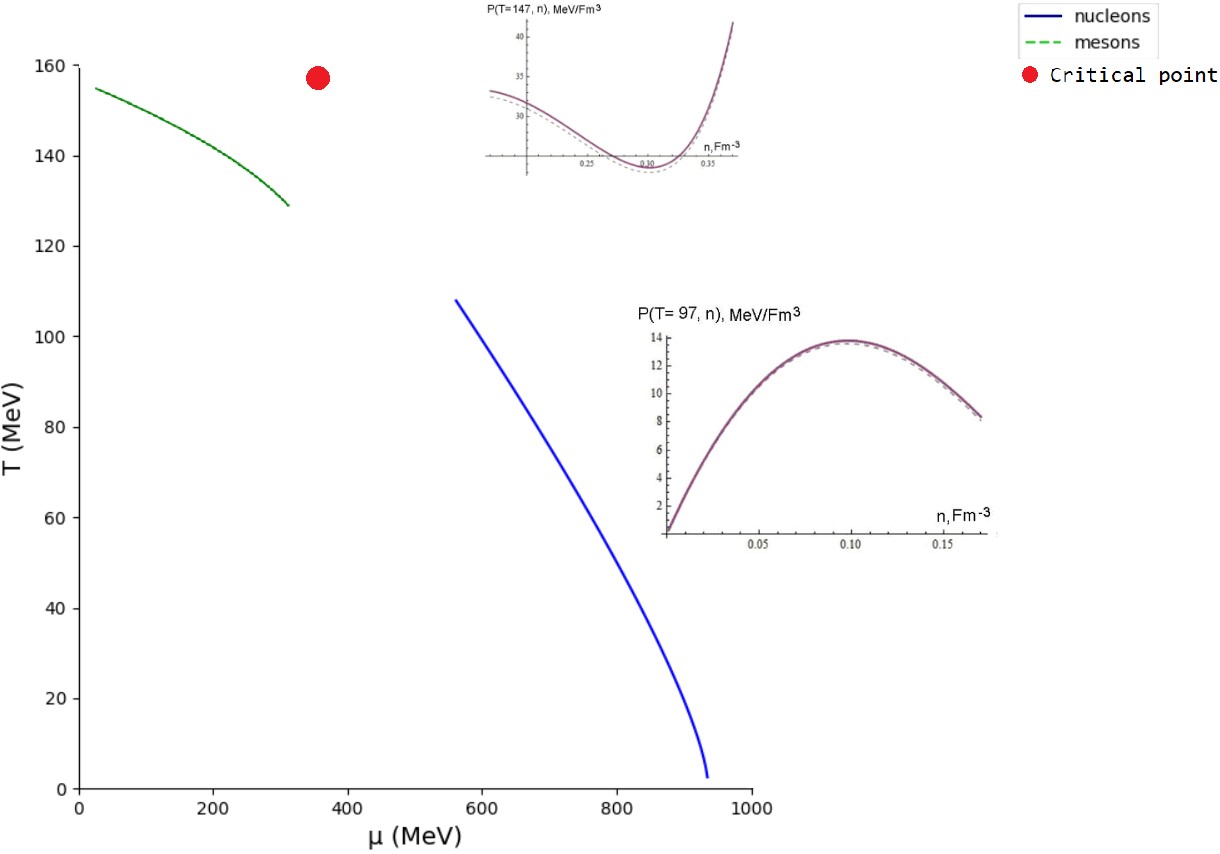}
\caption{The result of our calculations \cite{krivemet4} using formula (\ref{eq7771}) for the meson and nucleon stages of the evolution of the hadron fireball}
\label{pic7}
\end{center}
\end{figure}

In this approach, the system properties are determined by the pressure function  (\ref{eq:335}). The chemical potentials $\mu_i$ (where $i = 1, ..., K$) are defined as a combination of the baryonic $\mu_B$, strange $\mu_S$, and electric $\mu_Q$ chemical potentials, with coefficients of expansion $(\gamma_{B})_i$, $(\gamma_{S})_i$, and $(\gamma_{Q})_i$ respectively.


Interestingly, despite the crudeness of such a one-component approximation for the real multi-component vdW gas of the hadron fireball, as shown in Fig. \ref{pic7}, a good qualitative and quantitative agreement with the results of calculations by other authors is obtained for the chemical potential (see, for example,  \cite{stock1}, \cite{fukush} and  \cite{Karsch1}–\cite{Karsch2}).

\section{Summary}
The impact of considering the excluded volume and attraction is analyzed in the case of a two-component gas: (i) $\pi^0$- and $\pi^+$-mesons (model from Section 3); (ii) protons and neutrons (model from Section 4). The calculations were performed in the Canonical and Grand Canonical ensembles using the saddle point method for the two-component system. Particles interact with hard-core potentials at short distances and relatively high potentials at long distances (effective attraction radii). For such effective interparticle interactions, an equation of state with corrections that account for the finite dimensions of the nuclear fireball, as well as the root-mean-square fluctuations of pressure and density, has been derived. The pressure correction vanishes in the thermodynamic limit, in accordance with statistical physics, where there is no distinction between different statistical ensembles.

Formulas for pressure and density obtained through the saddle point method can be employed to analyze experimental data regarding the relative abundance of particles of different types and critical parameters in high-energy nuclear-nuclear collisions. As an example of such application for the chemical freeze-out stage (model from Section 3), a generalization of the presented van der Waals model to the case of an asymmetric two-component model ($\pi^0$- and $\pi^+$-mesons) with effective phenomenological hard-core and attractive parameters has been proposed. The ratio of the pressure correction to the root-mean-square value of pressure fluctuation is assessed for the case of an asymmetric two-component meson fireball model. An increase in the correction at low density values corresponding to the final freezing stages has been identified.

It has been found that the contribution to pressure, considering different radii and the finiteness of the nuclear fireball, in comparison to root-mean-square fluctuations, becomes noticeable in the case of the meson model with particle non-conservation (model from Section 3, corresponding to the chemical freeze-out stage). However, this correction can be neglected in the final stages of freeze-out when nucleons begin to dominate (model from Section 4, corresponding to the kinetic freeze-out stage).

The developed approach of integrating a large statistical sum using the saddle point method allows for obtaining both the equation of state and expressions for chemical potentials uniquely, and it can be easily extended to the case of a multi-component system (Section 5).

Interestingly, despite the simplicity of the single-component approximation, the obtained behavior of the baryon chemical potential qualitatively, and sometimes quantitatively, reproduces the corresponding calculations of other authors made under different QCD approximations.

Thus, it is anticipated that the developed model can be useful in analyzing experimental data related to the study of various stages of nuclear fireball evolution, which occurs, in particular, in experiments investigating the quark-gluon plasma.

The research was conducted within the framework of the initiative scientific topic 0122U200549 ("National Technical University of Ukraine 'Igor Sikorsky Kyiv Polytechnic Institute'").

\end{NoHyper}
\end{document}